\documentclass{aa}
\usepackage{natbib}
\usepackage{graphicx}
\usepackage{amsmath}
\usepackage{amssymb,amsmath,epsfig,array}
\usepackage{adjustbox}
\usepackage{multicol}
\usepackage{changes}

\def\b{${\beta}$}
\def\g{${\gamma}$}
\def\dP{$\dot{P}$}
\def\kms{km s$^{-1}$}
\def\Ms{$M_{\odot}$}
\def\Rs{$R_{\odot}$}
\def\Msy{$M_{\odot}$ yr$^{-1}$}

\title{Spectroscopic Searches for Evolutionary Orbital Period Changes in WR+OB Binaries: the Case of WR 127 (Hen 3-1772)}

\author{I.A. Shaposhnikov\inst{1,2}\thanks{iv.shaposhnikov@gmail.com}\,
\and
A.M. Cherepashchuk\inst{1}
\and
A.V. Dodin\inst{1}
\and
K.A. Postnov\inst{1,3}}

\institute{
Lomonosov Moscow State University, Sternberg Astronomical Institute, Universitetskij pr. 13, 119234 Moscow, Russia \and
Lomonosov Moscow State University, Faculty of Physics, Leninskiye Gory 1-2, 119991 Moscow, Russia \and
Kazan Federal University, Kremlyovskaya 18, 420008 Kazan, Russia
}
\date{Received 26 February 2024 / Accepted 4 March 2024}

\begin{document}

\abstract
{}
{
We aim to determine the secular evolution of the orbital period of the short-period binary system WR 127 (WN3b+O9.5V, $P \approx 9.555^d$).
}
{We performed new low-resolution spectroscopic observations of WR 127 with the 2.5 m CMO SAI telescope to construct the radial velocity curves of the components.\ Our results suggest component masses of $M_\mathrm{WR}\sin^3(i) = 11.8\pm1.4$ \Ms\ and  $M_\mathrm{O}\sin^3(i)=17.2\pm1.4$ \Ms.
By comparing these values with archival radial velocity curves we were able to create an $(O-C)$ plot with an accuracy sufficient to search for the orbital period change in WR 127.
}
{We report the reliable detection of a secular increase in the orbital period of WR 127 at a rate of $\dot{P} = 0.83\pm0.14~\mbox{s yr}^{-1}$, which corresponds to  a dynamical mass-loss rate from the Wolf-Rayet (WR) star of $\dot{M}_\mathrm{WR} = (2.6\pm0.5)\times 10^{-5}$ \Msy.} 
{The mass-loss rate from WR stars in three Wolf-Rayet+OB binaries (WR 127, CX Cep, and V444 Cyg) as inferred from spectroscopic and photometric measurements suggests a preliminary empirical correlation between a WR star's mass and its dynamical mass-loss rate of $\dot M_\mathrm{WR}\sim M_\mathrm{WR}^{1.8}$.
This relation is important for the understanding of the evolution of massive close binaries that include WR stars as such an evolution is a precursor of gravitational-wave binary merging events with neutron stars and black holes.
} 


\titlerunning{Orbital Period Changes in WR+OB Binaries (WR 127)}
\authorrunning{I.A. Shaposhnikov et al.}
\maketitle

\section{Introduction}

Determining the mass-loss rate of stellar winds from Wolf-Rayet (WR) stars is crucial to understanding their evolution. The most reliable estimates of WR mass-loss rates can be obtained from measurements of the secular evolution of the orbital periods of WR stars in close binaries. Such measurements are easiest for the numerous close WR+OB binaries with early-type O or B secondary components. The orbital period evolution can be inferred from the analysis of an $(O-C)$ plot calculated for moments of orbital eclipses. The primary eclipses are best computed by overlaying orbital light curves obtained at different epochs, the so-called modified Hertzsprung method. Unfortunately, the number of WR+OB binaries with pronounced eclipses is rather small, which significantly lowers the prospects of such a study. 

The system WR 127 = HD 186943 = Henize 3-1772 is an example of a WR+OB close binary that does not experience deep photometric eclipses. In a past study, using the eclipsing WR+OB binaries  CQ Cep, CX Cep, and V444 Cyg, we demonstrated that the orbital period change rate, \dP, as inferred from a comparison of radial velocity curves obtained at different epochs, is consistent with estimates obtained using the modified Hertzsprung method \citep{Shaposhnikov2023a,Shaposhnikov2023b}. Thus, for close binaries with WR stars for which the photometric behaviour does not enable the application of the classical Hertzsprung method to search for orbital period variations, there is the possibility to estimate \dP\ by comparing radial velocity curves, provided that there is a  sufficient number of archive radial velocity measurements. Most close binaries containing WR stars in the northern sky are well studied; the first radial velocity measurements were carried out as early as the 1940s. 

In the present paper we report on searches for the orbital period increase in WR 127 and provide a dynamical estimate of the WR mass-loss rate from measurements of WR 127 radial velocity curves. 

\section{Observations}

\subsection{Archival data}

\begin{figure*}[h!]
  \centering
  \includegraphics[scale=0.55]{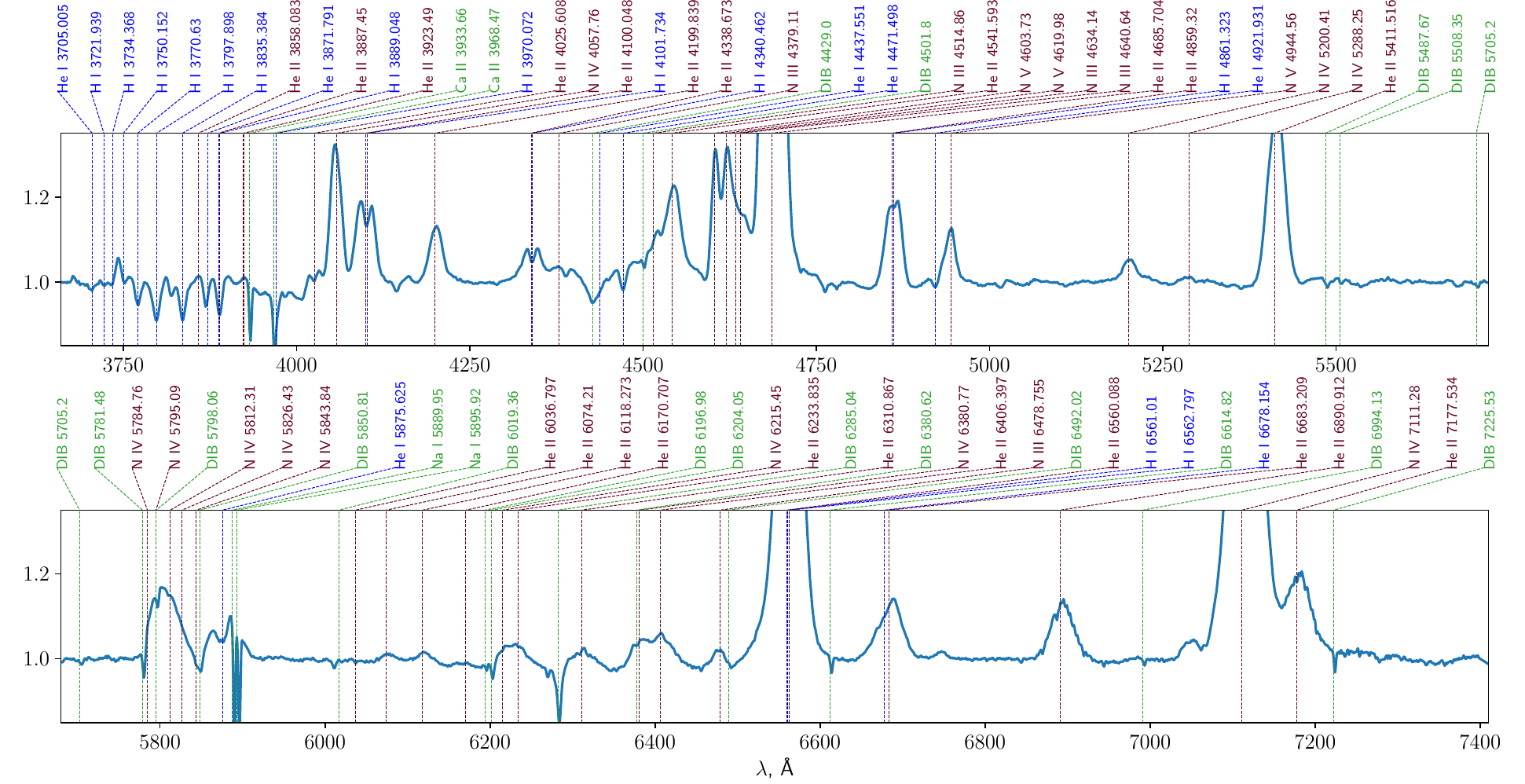}
  \caption{Average 
  spectrum of WR 127 from our observations (Table 1). WR emission lines are in dark red, the O-star absorptions are in blue, and interstellar features are in green.}
 \label{mean}
\end{figure*}

The spectral binarity of the WR star HD 186943 was reported by \citet{Wilson1941}. 
\cite{Hiltner1945} carried out the first detailed spectroscopic study of this system and presented radial velocities of the WR star (classified by Hiltner as WN5) in the emission lines \ion{He}{II} (4686 \AA) and \ion{N}{V} (the mean of the 4603 \AA\ and 4619 \AA\ line components) and in hydrogen absorptions (the mean of H\b, H\g, and H8-11). 

The HD 186943 radial velocity measurements in the emission lines \ion{He}{II} 4686 \AA, \ion{N}{V} 4603 \AA, \ion{and N}{IV} 4058 \AA\ obtained from ten spectrograms were published by \citet{Ganesh1967}. More detailed radial velocity curves and tabulated measurements in the same emission lines and absorption lines were presented by \citet{Massey1981}. 
He classified the system as WN4+O9V and estimated the orbital elements; based on accepted estimates of the spectral type, O-star mass, and distance, he estimated that the inclination was large and predicted that the system would eclipse.

\citet{Hamann2006} conducted extensive research to determine the parameters of a large number of WR stars using Potsdam Wolf-Rayet models  that include line-blanketing \citep{Grafener2002}. WR 127 was excluded from the main study due to its composite spectrum, but he proposed the spectral classification WN3+O9.5V.

\citet{Chevrotiere2011} presented an analysis of 23 spectra of WR 127. The authors classified the binary as WN5o+O8.5V and reported radial velocity measurements in \ion{N}{V} emissions (4603 \AA, 4619 \AA, and 4945 \AA) and  \ion{H}{I} absorptions (3835 \AA, 3798 \AA, and 3771 \AA). In addition to improving the orbital system parameters, the authors also studied parameters of the interaction zone of the components' stellar winds.

\citet{Dsilva2022} conducted a spectroscopic survey of several known WR nitrogen stars to study their spectral binarity. Observations were carried out with the High-Efficiency and high-Resolution Mercator Echelle Spectrograph (HERMES).  
%\LEt{ Consider defining. ***}. 
The radial velocities were measured by cross-correlating the \ion{N}{V} 4945 \AA\ line. The authors present 11 radial velocity curves for WR 127, among others. 

There are also archival photometric observations of WR 127. The binary inclination angle estimated by Massey and the possibility to observe eclipses in this system motivated various authors to perform photometric observations of WR 127

Observations by \citet{Lipunova1984,Lipunova1986} revealed a weak brightness variability consistent with the orbital period; however, the reliability of this result was questioned by the author due to a bad choice of comparison star. A broad but shallow minimum (about 0.03 mag) in the phasing curves was noted by \citet{Moffat1986}, and \citet{Lamontagne1996} found the solution of this light curve. In the archives, one can find optical light curves obtained from space observations by \textit{Hipparcos} \citep{Marchenko1998b} and the INTErnational Gamma-Ray Astrophysics Laboratory (\textit{INTEGRAL}) satellites
\citep{Alfonso2012}. A number of brightness measurements of WR 127 can also be found in the  All Sky Automated Survey for SuperNovae (ASAS-SN) database \citep{Shappee2014,Kochanek2017}.

\begin{figure}[h]
 \centering
 \includegraphics[scale=0.65]{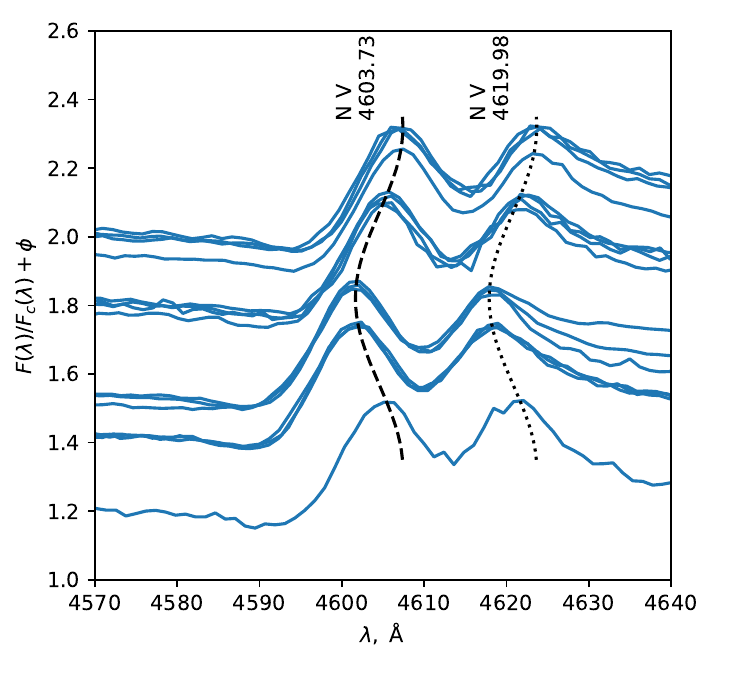}
  \caption{WR 127 spectra from Table 1 near the \ion{N}{V} 4603 \AA\ line, shifted by orbital phase.}
 \label{sp_nv}
\end{figure}

\subsection{New observations}

We carried out the latest spectral observations of WR 127  between October 2022 and January 2024 using the 2.5 m telescope at the Caucasus Mountain Observatory of Sternberg Astronomical Institute (CMO SAI) 
using the low-resolution ($R\sim 1500-2000$) Transient Double-beam Spectrograph (TDS) 
\citep{Potanin2020}. The observations and data reduction are described in detail in \citet{Shaposhnikov2023a,Shaposhnikov2023b}. 

\begin{figure}[h]
 \centering
 \includegraphics[scale=0.56]{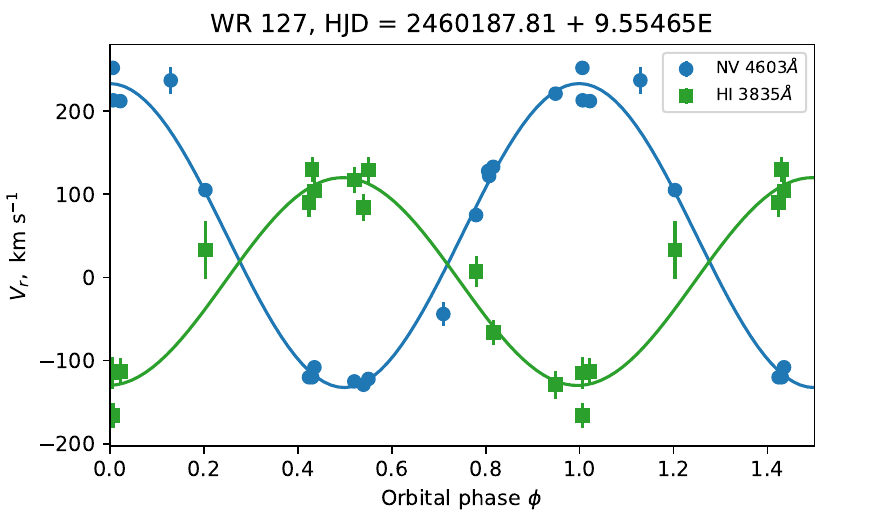}
  \caption{Radial velocity curves of WR 127 in the \ion{N}{V} 4603 \AA\ and \ion{H}{I} 3835 \AA\ lines.}
 \label{rc}
\end{figure}

Figure \ref{mean} shows the spectrum of WR 127 averaged over all our observations without any radial velocity shifts. Figure \ref{sp_nv} shows  the spectra listed in Table 1  near the \ion{N}{V} 4603 \AA\ line, which we used to plot the $(O-C)$ diagram. Figure \ref{rc} presents the radial velocity curves of the WR and O-star in this system in \ion{N}{V} 4603 \AA\ and \ion{H}{I} 3835 \AA\ lines, respectively. 
The radial velocity semi-amplitudes are $K_{\mathrm{WR}} = 182.7\pm2.0$ \kms\ and $K_{\mathrm{O}} = 125\pm5$ \kms\ (adopting a circular orbit). 
The corresponding component masses and orbital radii are $M_\mathrm{WR}\sin^3(i) = 11.8\pm 1.4$ \Ms,  $M_\mathrm{O}\sin^3(i)=17.2\pm 1.4$ \Ms, and 
$a_\mathrm{WR}\sin(i) =34.6\pm0.4$ \Rs, $a_\mathrm{O}\sin(i)=23.6\pm2.1$ \Rs, respectively.
Radial velocity measurements for each night are presented in Table \ref{vels}. These data enabled us to improve the system's orbital elements, which, along with the estimates obtained by other authors, are summarized in Table \ref{comparams}. 

\begin{table}[h!]
    \centering
    \scriptsize
    \caption{Radial velocity measurements of WR 127 from the CMO SAI observations (\kms).}
    \begin{tabular}{cccc}
\hline\hline
HJD       &  V(\ion{N}{V} 4603\AA) &  V(\ion{N}{V} 4619\AA)  &  V(\ion{N}{V} 4945\AA)  \\
\hline
2459855.33738 &  105$\pm$7  &   85$\pm$11 &   50$\pm$13 \\
2459883.29626 &  237$\pm$16 &  224$\pm$24 &  151$\pm$11 \\
2459906.14007 & -125$\pm$6  & -107$\pm$6  & -194$\pm$7  \\
2460029.53924 & -108$\pm$6  &  -67$\pm$9  & -166$\pm$7  \\
2460061.49213 &   75$\pm$6  &  126$\pm$8  &   45$\pm$10 \\
2460178.31632 &  213$\pm$8  &  243$\pm$9  &  137$\pm$6  \\
2460182.30322 & -120$\pm$5  &  -88$\pm$10 & -203$\pm$9  \\
2460183.41227 & -129$\pm$6  &  -99$\pm$8  & -195$\pm$6  \\
2460243.29696 &  122$\pm$5  &  167$\pm$9  &   76$\pm$12 \\
2460245.19439 &  252$\pm$6  &  308$\pm$15 &  171$\pm$7  \\
2460290.13792 &  -44$\pm$14 &      --     &      --     \\
2460291.15237 &  133$\pm$7  &   189$\pm$9 &   50$\pm$8  \\
2460293.11999 &  212$\pm$6  &   277$\pm$9 &      --     \\
2460298.16879 & -122$\pm$6  &  -101$\pm$7 & -202$\pm$8  \\
2460310.15951 &  128$\pm$7  &      --     &      --     \\
2460316.13260 & -120$\pm$7  &  -51$\pm$9  & -189$\pm$9  \\
\hline
\\
\hline\hline
HJD         &  V(\ion{H}{I} 3770\AA) & V(\ion{H}{I} 3798\AA)  &  V(\ion{H}{I} 3835\AA)  \\
\hline
2459855.33738 &  -51$\pm$32  &  47$\pm$26  &   33$\pm$35  \\
2459883.29626 &       --     &     --      &       --     \\
2459906.14007 &   71$\pm$17  &  21$\pm$21  &  117$\pm$16  \\
2460029.53924 &  114$\pm$24  &  53$\pm$19  &  104$\pm$19  \\
2460061.49213 &   53$\pm$32  & -65$\pm$18  &    7$\pm$19  \\
2460178.31632 &  -73$\pm$22  & -211$\pm$14 & -166$\pm$15  \\
2460182.30322 &  183$\pm$19  &  78$\pm$16  &  90$\pm$17   \\
2460183.41227 &  160$\pm$16  &  51$\pm$14  &  84$\pm$16   \\
2460243.29696 &       --     &      --     &      --      \\
2460245.19439 &       --     & -100$\pm$17 & -115$\pm$19  \\
2460290.13792 &       --     &      --     &      --      \\
2460291.15237 &   -4$\pm$19  & -111$\pm$16 &  -66$\pm$15  \\
2460293.11999 &  -49$\pm$22  & -128$\pm$21 & -113$\pm$16  \\
2460298.16879 &   91$\pm$19  &   28$\pm$16 &  129$\pm$16  \\
2460310.15951 &       --     &     --      &     --       \\
2460316.13260 &       --     &   32$\pm$20 &  130$\pm$15  \\
\hline
    \end{tabular}
    \label{vels}
\end{table}

\section{Dynamical estimate of the orbital period change, \dP, and the WR mass-loss rate, $\dot{M}_{\mathrm{WR}}$}

\begin{table*}
    \centering
    \scriptsize
    \label{comparams}
    \caption{Parameters of WR 127 from different papers. }   

\begin{tabular}{cccccccc}
\hline
\hline
Ref.    & Mean HJD & SpT   & $P$, days & $K_\mathrm{WR}$(\ion{N}{V}), \kms & $K_\mathrm{O}$, \kms & $M_\mathrm{WR}\sin^3(i)$, \Ms & $M_\mathrm{O}\sin^3(i)$, \Ms \\
\hline
H45   & 2431281  & WN5+early B & 9.550  & 160$\pm$10\tablefootmark{a} & $\sim$65       & $\sim$5.8         & $\sim$21     \\
GB67   & 2434201  & WN5, WN5-A  & 9.5594 & $\sim$162.5 & --       & --  & --        \\
M81  & 2444090  &  WN4+O      &  9.555        & 178$\pm$7 &  99$\pm$3  & $\sim$9          & $\sim$20        \\
Ch11  & 2453195 & WN5o+O8.5V  & 9.555 & 177$\pm$5     & 99$\pm$4   & 7.5$\pm$0.6  & 13.4$\pm$1.0  \\
D22  & 2458256  & WN3b+O9.5V  & 9.555 & 163$\pm$5  & --        & --  & --   \\
Our data & 2460170 & WN3b+O9.5V  & 9.55465 & 182.7$\pm$2.0 & 125$\pm$5 & 11.8$\pm$1.4  & 17.2$\pm$1.4    \\
\hline
\end{tabular}
\\

    \begin{tabular}{cccccc}
\\
\hline
\hline
Ref.    &  Type & $i$, $^{\circ}$ & M$_\mathrm{WR}$, \Ms & M$_\mathrm{O}$, \Ms & $\dot{M}_\mathrm{WR}$, $10^{-5}$\Msy \\ 
\hline
\citet{StLouis1988}  & Pol.  & 55.7$\pm$8.2 & 16$\pm$5 & 35$\pm$11 & $\sim$0.8        \\
\citet{Lamontagne1996} & Phot. & 55.3$\pm$4.7 & $\sim$17       & $\sim$36        & 1.4$\pm$0.4\tablefootmark{b} \\
Our data (2022-2024) & Spec. & 55.5 & 21.0$\pm$2.5 & 30.8$\pm$2.4 & 2.6$\pm$0.5 \\
\hline
    \end{tabular}
\tablefoot{
The upper table contains the results of spectroscopic observations of WR 127 obtained in different years. 
The lower table contains the observational results from which estimates of orbital inclination $i$ and/or mass loss $\dot{M}_{\mathrm{WR}}$ rate are derived. \\
\tablefoottext{a}{
calculated by us from radial velocity curve for \ion{N}{V} by \citet{Hiltner1945};}
\tablefoottext{b}{based on the optical eclipse depth.}
}
\tablebib{
(H45) \citet{Hiltner1945}; (GB67) \citet{Ganesh1967}; (M81) \citet{Massey1981}; (Ch11) \citet{Chevrotiere2011}; (D22) \citet{Dsilva2022}
}

\end{table*}

\begin{figure}[h]
 \centering
 \includegraphics[scale=0.55]{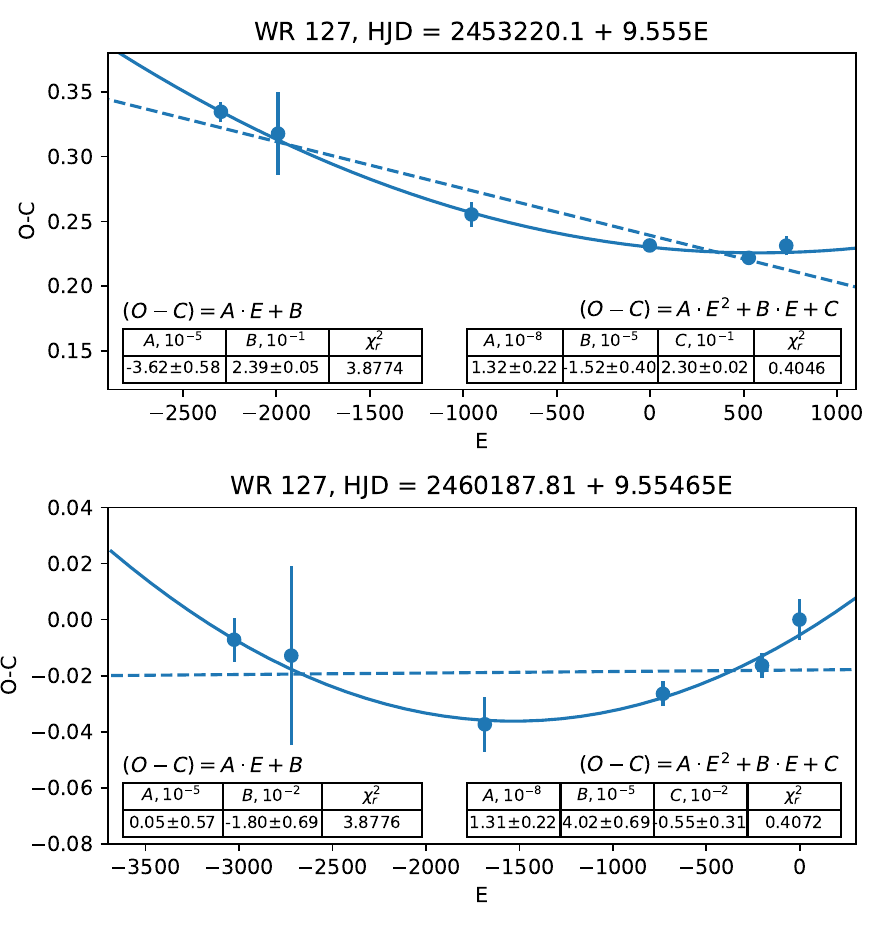}
  \caption{$(O-C)$ plot constructed from radial velocity curves of WR 127 and tables with coefficients of linear and quadratic approximations. The top and bottom panels show plots constructed using the ephemeris from \citet{Chevrotiere2011} and the improved ephemeris from our data, respectively.}
 \label{oc}
\end{figure}

Figure \ref{oc} shows $(O-C)$ plots of spectroscopic observations of WR 127. The photometric light curves of WR 127 are noisy and can be collected over time intervals shorter than 30 years. Their comparison does not allow us to reliably conclude on the orbital period evolution. In contrast, our comparison of radial velocity curves clearly indicates an increase in the orbital period: the parabolic arms in Fig. \ref{oc} are directed upwards. As in the case with V444 Cyg \citep{Shaposhnikov2023b}, here we have almost exclusively used radial velocity curves 
in the \ion{N}{V} 4603 \AA\ line; the only exception is high-resolution measurements in the \ion{N}{V} 4945 \AA\ line taken from \citet{Dsilva2022}.

The upper panel of Fig. \ref{oc} presents the difference between the radial velocity maximum phase and the conjunction phase calculated using the $P$ and $E_0$ from the ephemeris from \citet{Chevrotiere2011}: $P=9.555^d$ and $E_0 = 2453220.1$.
The strong inclination in the plot suggests an incorrect value for the adopted period, $P$. We also see that a parabola clearly provides a much better fit than a straight line.

The $(O-C)$ plot in the bottom panel of Fig. \ref{oc} was constructed using a modified orbital ephemeris. The epoch $E_0 = HJD~2460187.81^d \pm 0.15$ occurred at the radial velocity maximum determined from our own observations, and the orbital period $P=9.55465\pm0.00005^d$ was chosen to remove the linear coefficient from the approximation. 

The tables in Fig. \ref{oc} show the linear and quadratic fit coefficients and the reduced $\chi^2_r$ values. The null hypothesis is the linear fit to the $(O-C)$ plot, which corresponds to a constant orbital period, $P$. A quadratic fit corresponds to the tested hypothesis regarding a non-zero orbital period change rate, \dP. The $\chi^2_r$ value is  3.88 and 0.41 for the linear and quadratic fit, respectively, refuting the null hypothesis in favour of actual change in the orbital period of WR 127 (an increase since $A>0$). Although the $\chi_r^2$ for the quadratic fit is notably smaller than 1, we do not think that this indicates an uncertainty in the model but is instead due to the parabola passing very close to the $(O-C)$ points.
Based on the representation of the quadratic coefficient $\dot{P} = 2\times  A,$ we can calculate the orbital period increase rate of WR 127:   $\dot{P} = (2.6\pm0.4) \cdot 10^{-8} ~\mathrm{s~s}^{-1} = (2.5\pm0.4) \cdot 10^{-7} ~\mathrm{days~per~orbit} = 0.83\pm0.14$ s yr$^{-1}$. 
Thus, the WR radial velocity maxima are
\begin{eqnarray}
    &T(\mathrm{max}~V_\mathrm{WR}) = E_0 + P \times E + 0.5\dot{P}\times  E^2 = \nonumber \\
    &=
    \begin{matrix}
        2460187.81 & + 9.55465 E & + 1.25\cdot10^{-7} E^2\,.\\
        \pm0.15    & \pm0.00005  & \pm0.21\cdot10^{-7}
    \end{matrix}\nonumber
\end{eqnarray}

Applying the Jeans mode of the mass outflow from the binary and neglecting the mass loss from the O9.5V star, we find $\dot{M}_{\mathrm{WR}} = -\frac{1}{2} \frac{\dot{P}}{P} (M_{\mathrm{WR}}+M_{\mathrm{O}})$. By adopting the average value of the binary inclination angle derived from the photometric \cite{Lamontagne1996} and polarimetric \cite{StLouis1988} observations, $i = 55.5^{\circ}$, and using the $M\sin^3i $ from our analysis (see Table 2), we estimate the masses of the components to be $M_{\mathrm{WR}}=21.0\pm2.5$ \Ms\ and $M_{\mathrm{O}}=30.8\pm2.4$ \Ms.
Using the component mass estimates and the derived value of the orbital period increase rate,  $\dot{P} = 0.83\pm0.14$ s yr$^{-1}$, we finally obtain the mass-loss rate of
the WR star: $\dot{M}_{\mathrm{WR}} = (2.6\pm0.5)\times 10^{-5}$ \Msy.

We note that accounting for finite stellar radii when calculating $\dot{M}_{\mathrm{WR}}$, which can be appropriate for the very close binaries CQ Cep, CX Cep, and V444 Cyg \citep{Shaposhnikov2023a,Shaposhnikov2023b}, appears unnecessary for WR 127. Even for V444 Cyg, whose orbital period is more than twice shorter than that of WR 127 ($P \approx 4.21^d$), accounting for the finite stellar radii only leads to a $3\%$ correction in the dynamical estimate of $\dot{M}_{\mathrm{WR}}$ \citep{Shaposhnikov2023b}.

\section{Conclusion}
New low-resolution spectroscopic measurements 
of WR 127 (WN3b+O9.5V) were carried out in 2022-2024 with the 2.5 m telescope CMO SAI. Using these observations and archival radial velocity curve measurements, we constructed a spectroscopic $(O-C)$ plot, which enabled us to determine that the secular orbital period of the binary increases at a rate of $\dot{P} = 0.83\pm0.14$ s yr$^{-1}$. 
Assuming a Jeans mode mass outflow from the system, we estimated the WN3b  stellar wind mass-loss rate to be  $\dot{M}_\mathrm{WN3b} = (2.6\pm0.5)\times 10^{-5}$ \Msy.

\begin{figure}[h]
 \centering
 \includegraphics[scale=0.55]{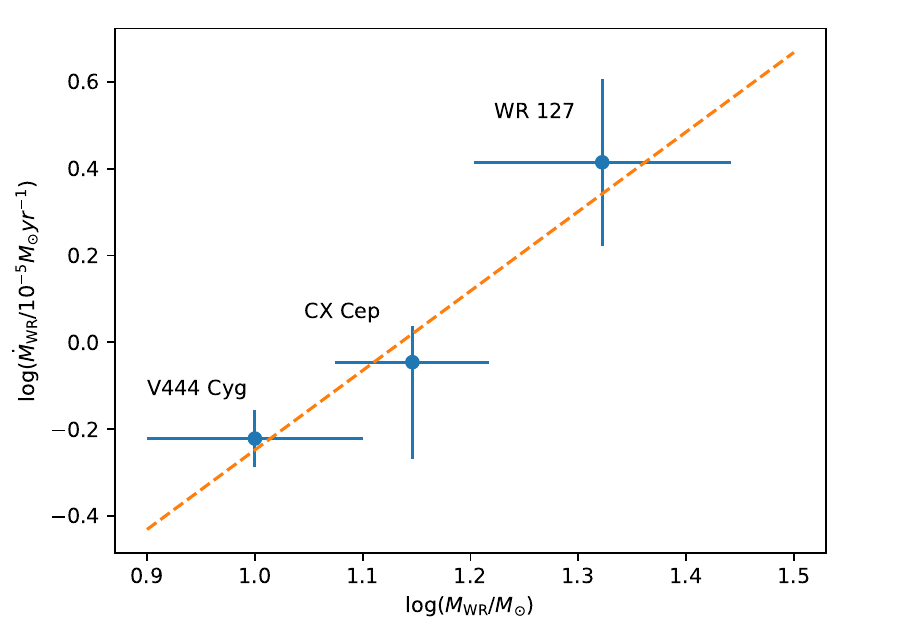}
  \caption{Empirical correlation between the mass and mass-loss rate of WN stars determined via the spectroscopic methods from \cite{Shaposhnikov2023a,Shaposhnikov2023b} and the present paper.}
 \label{dm-m}
\end{figure}

We emphasize that in our case for the WN3b star with mass  $M_{\mathrm{WN3b}} \approx 21$ \Ms, the determined mass-loss rate is  $\dot{M}_{\mathrm{WN3b}} \approx 2.6\times 10^{-5}$ \Msy.\ For comparison, the mass-loss rate for the WN5 star in CX Cep with $M_{\mathrm{WN5}} \approx 14$ \Ms\  is $\dot{M}_{\mathrm{WN5}} \approx 0.9\times 10^{-5}$ \Msy,  and for the WN5 star with mass  $M_{\mathrm{WN5}} \approx 10$ \Ms\ in V444 Cyg it  is $\dot{M}_{\mathrm{WN5}} \approx 0.6\times 10^{-5}$ \Msy\ \citep{Shaposhnikov2023a,Shaposhnikov2023b}.

This suggests an empirical correlation between the WR star mass and the reliable dynamical mass-loss rate estimate (see Fig. \ref{dm-m}): 
\begin{equation}
    \log\left(\frac{\dot M_{\mathrm{WR}}}{10^{-5} M_\odot \mathrm{yr}^{-1}}\right)= (1.8\pm0.4)\log\left(\frac{M_{\mathrm{WR}}}{M_\odot}\right)-(2.1\pm0.5).\nonumber
\end{equation}

Making use of the spectral method elaborated in  \cite{Shaposhnikov2023a,Shaposhnikov2023b} and in the present paper, we hope to provide dynamical mass-loss estimates for other spectroscopic double WR+OB systems, with the aim of determining the  empirical $\dot M_{\mathrm{WR}} - M_{\mathrm{WR}}$ relation.
New observations will allow us to improve the $\dot M_{\mathrm{WR}} - M_{\mathrm{WR}}$ correlation suggested from the analysis of the linear polarization of WR+OB binaries \citep{StLouis1988}. Such a correlation is very important towards understanding the evolution of close massive binaries, which are progenitors of coalescing double neutron stars and black holes. 

\begin{acknowledgements}

The work of IASh, AMCh and AVD (observations, data reduction and analysis of spectral observations) is supported by the Russian Science Foundation through grant 23-12-00092. KAP acknowledges the collaboration with 
the International Space Science Institute (ISSI) in Bern, through ISSI International Team project 512 "Multiwavelength view on massive stars in the era of multimessenger astronomy".

\end{acknowledgements}

\bibliographystyle{aa}
\bibliography{bib}
\end{document}